\documentclass[twocolumn,pra,superscriptaddress]{revtex4-2}

\usepackage{epsfig}
\usepackage{graphicx}
\usepackage{amsmath}
\usepackage{amssymb}
\usepackage{epstopdf}
\usepackage{xcolor}
\usepackage[caption=false]{subfig}
\usepackage{bm}
\usepackage{dsfont}
\usepackage{hyperref}
\usepackage{braket}
\usepackage{todonotes}
\usepackage{comment}

\graphicspath{ {./Images_state_tomography/} }

\begin{document}

\title{Efficient full state tomography of the orbital angular momentum states of light using Helicity sorter}
\author{Joel M.~Sunil}
\email{joel.sunil@gmail.com}
\affiliation{Department of Physics, BITS Pilani K K Birla Goa Campus, 
 NH17B, Zuarinagar, Goa 403726, India}

\author{Sandeep K.~Goyal }
\email{skgoyal@iisermohali.ac.in}
\affiliation{Department of Physical Sciences, Indian Institute of Science Education \& Research (IISER) Mohali,
  Sector 81 SAS Nagar, Manauli PO 140306, Punjab, India.}

\begin{abstract}
Orbital angular momentum (OAM) of light is a promising degree of freedom for quantum communication and quantum computation. However, efficient methods to transform and measure OAM modes are still not available readily.  Here, we propose a method for full quantum state tomography for arbitrary OAM states of light. The key component in our scheme is the helicity sorter, a device that can sort OAM modes of a light beam based on their  Helicity. We present two setups to implement the Helicity sorted, one requires $\mathcal{O}( N)$ number of optical elements for $2^N$ OAM modes, and the other uses spatial light modulators and is independent of the number of OAM modes.  Ours is the first method to provide the efficient and scalable full quantum state tomography of the OAM modes and can be extremely important for quantum communication and quantum computation. 
\end{abstract}

\maketitle

\section{Introduction}
Quantum information processing (QIP) offers a new avenue for efficient computation and most secure communication. QIP tasks require quantum systems, which can be efficiently manipulated, measured and are robust against environmental effects. Orbital angular momentum (OAM) states of light is among the most prominent  candidates for QIP tasks \cite{Allen1992,Vanenk1994,Yao2011,Molina2007,Cozzolino20192,Cozzolino2019,Ambrosio2012}. One of the biggest difficulties in working with OAM states of light is performing the quantum measurements  and state tomography on these states. There are several schemes to perform measurements on the OAM states of light~\cite{Leach2002,Leach2004,Berkhout2014,Mirhosseini2013,Sahu2018,Fickler2016,Kulkarni2017,DiLorenzoPires2010,Ghai2009,Christina2004,Sztul2006,Ghai2009,Yongxin2011,Anderson2012,Liu2019}, but all of them have their limitations. % One of the reason for the inefficiency of the measurements on the OAM modes is the inability to perform arbitrary unitary transformations on these systems.
Here, we propose a method to perform full quantum state tomography (QST) on an arbitrary state of a light beam. 

% OAM states are eigenmodes  of paraxial light beams which carry orbital angular momentum in addition to the spin angular momentum associated with polarization \cite{Allen1992,Vanenk1994}. They are characterized by helical wavefronts and a winding number or topological charge $\ell$ which can take any integer value, offering higher capacity to carry information. 

QST estimates the density matrix of a system by performing a complete set of measurements on an ensemble of identically prepared systems. Typically, QST protocol require at least $N+1$ mutually independent measurement operators in order to estimate the density operator of $N$-dimensional system~\cite{Adamson2008,Filippov2011,Yan2010}. The measurement operators whose eigenbases are mutually unbiased are the best choice for QST protocol~\cite{Wooters1989,Giovannini2013,Adamson2021,Czerwinski2021}. However, for large dimensions $N$, performing measurements in $N+1$ mutually independent operators can be a difficult task.

QST on the OAM states of light is especially challenging \cite{Agnew2011,Langford2004,Giovannini2013}. % \Red{First of all, it is difficult to prepare setups that are capable of performing $N+1$ independent measurement operators. In such cases, one is content with performing the measurement in a specific basis, such as the OAM basis.}
Often the researchers are content with measuring the population of each of the OAM modes which amounts to partial state tomography. This can be achieved by sorting the OAM modes in distinct paths and performing intensity measurement in each path. This method requires cumbersome interferometric setups and are nonscalable~\cite{Leach2002,Leach2004}. OAM modes can also be measured by performing a log-polar transformation that converts the helical phase into a linear phase~\cite{Berkhout2014,Mirhosseini2013,Sahu2018}.

QST on the OAM states is further made difficult by the fact that unitary transformations that are required to perform measurements in the mutually unbiased bases are very resource intensive. Methods to perform only SU(2) transformations efficiently are known on specific  OAM subspaces~\cite{gao,Padgett1999}. % As a result, QST on the OAM states requires many more measurements than the $N+1$ required with MUBs ~\cite{Agnew2011}.

There are alternate ways to perform measurements on the OAM modes that include schemes by measuring the transverse intensity profile of the light beams~\cite{Fickler2016,Kulkarni2017,DiLorenzoPires2010,Ghai2009,Christina2004,Sztul2006,Ghai2009,Yongxin2011,Anderson2012,Liu2019}. The Auxilliary Hilbert Space Tomography (AHST) proposed by Liu et.~al~\cite{Liu2019}  is one such scheme that performs a partial QST on the OAM states.

In this paper, we present a method to perform a full QST on the OAM states. The key component of our scheme is a device named helicity sorter. As the name suggests, the helicity sorter can sort the OAM states based on the helicity. This allows us to perform operations on negative and positive subspaces of the OAM modes independently and SU($2$) transformations between these two subspaces. Here we propose two schemes to realize helicity sorter, one uses only linear optical setup and requires $O(N)$ number of optical elements to sort $2^N$ OAM modes, whereas other scheme uses SLMs and is independent of the number $N$. The proposed QST scheme uses the AHST method of partial QST along with the Helicity sort.

Ours is the first method to perform a full QST on OAM states of light. Another highlight of our scheme is that it requires only three measurement settings for any arbitrary number of OAM modes. The linear optical nature and the finite number of measurement settings make our scheme efficient and scalable, and can be a turning point in the OAM based quantum computation and communication.

The paper is organised as follows: in Sec.~\ref{Sec:Background}~ we present the relevant background needed for our work. In Sec.~\ref{Sec:Results} we present the scheme for the full QST and the implementation schemes for the Helicity sorter. We conclude in Sec.~\ref{Sec:Conclusion}.

\section{Background}\label{Sec:Background}
In this section, we first introduce two linear optical devices that are essential to realizing a helicity-sorter: the OAM-sorter and the Gouy phase radial mode sorter. We also summarize the AHST scheme for QST~\cite{Liu2019}.
\subsection{The OAM-sorter}

An OAM sorter is a Mach-Zehnder interferometer that can sort OAM modes based on their $\ell$ values~\cite{Leach2002, Leach2004, Goyal2021}. For example, the device can be tuned in such a way that the even and odd $\ell$ modes will be sorted in separate paths. 
For an OAM sorter we require a linear optical device called Dove prism (DP). The action $\mathcal{D}(\beta)$ of a DP rotated by an angle $\beta$ about the axis of propagation of light on  OAM modes can be written 
\begin{align}
  \mathcal{D}(\beta) \ket{\ell} = e^{i2\ell\beta}\ket{-\ell}.
\end{align}

\begin{figure}
    \includegraphics[width=0.4\textwidth]{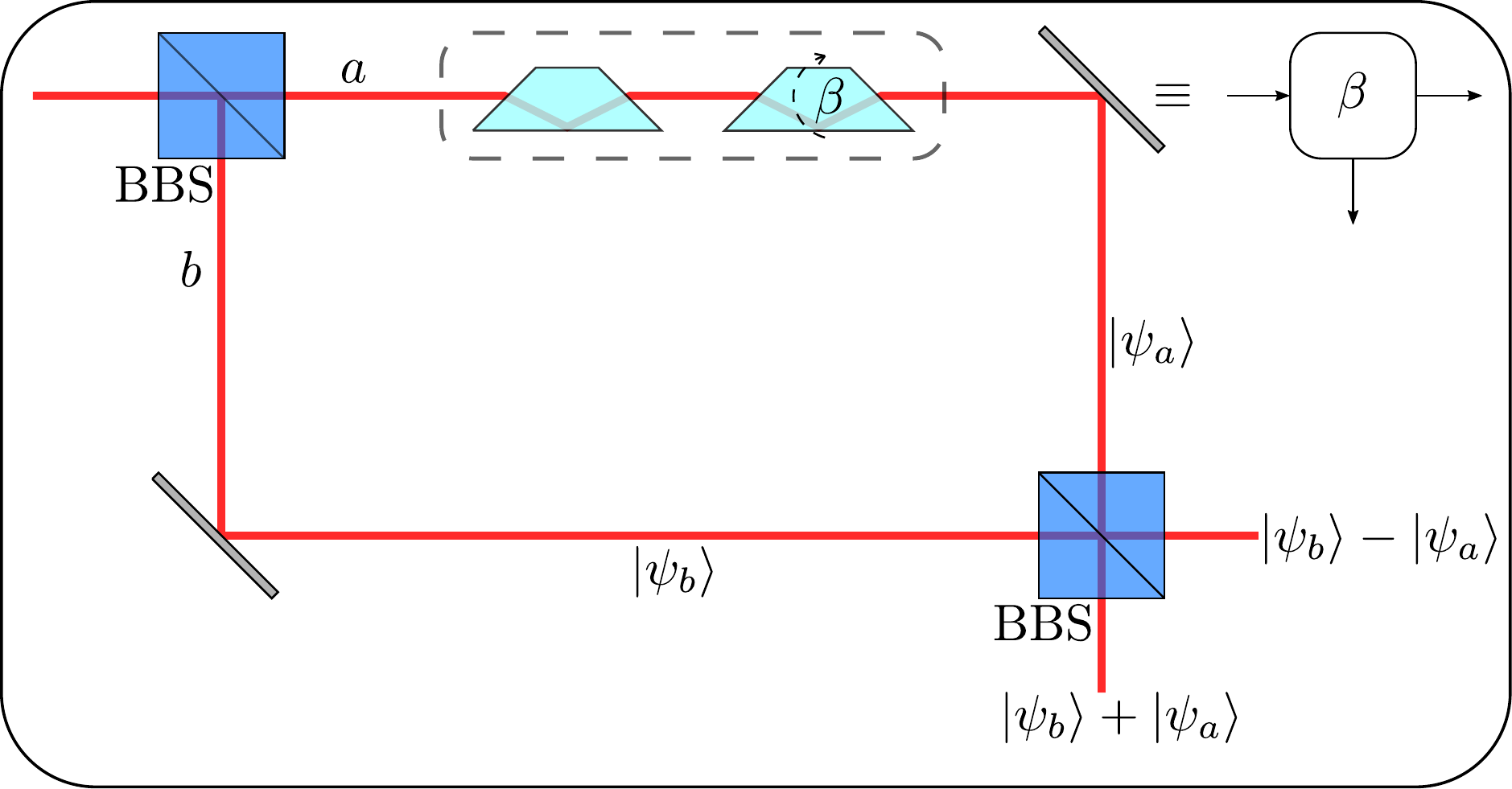}
  \caption{The OAM-sorter. Here we have a Mach-Zehnder interferometer with two Balanced Beam Splitter (BBS). Two DPs are placed in the path $a$ of the interferometer which attribute an $\ell$ dependent phase to the OAM modes.}
  \label{fig:OAM-sorter}
\end{figure}

In an OAM-sorter  two DPs are introduced in one arm of a Mach-Zehnder interferometer (Fig.~\ref{fig:OAM-sorter}). One of the DP is rotated at an angle $\beta$ and the other is kept at zero angle. This setup of DPs will leave the OAM mode unaffected and impart a phase $e^{i2\ell\beta}$ to the mode.

% Throughout this paper we will use the notation on the right side of fig. \ref{fig:OAM-sorter} to denote an OAM-sorter for compactness.
If a  state $\ket{\psi} =\sum_\ell g_\ell\ket{\ell}$ enters the Mach-Zehnder setup, then in path $a$ and path $b$ the states read
\begin{align}
\ket{\psi_a}&=\sum_\ell g_\ell e^{i2\ell\beta}\ket{\ell}\nonumber\\
\ket{\psi_b}&=\sum_\ell g_\ell \ket{\ell}.
\label{OAMsorter}
\end{align}
At the output of the interferometer the states reads $(\ket{\psi_a} \pm \ket{\psi_b})/\sqrt{2}$. Therefore, choosing  $\beta=\frac{\pi}{2}$ only either even or odd modes will survive in the output modes. Hence, one can sort the even and odd modes. The same setup can be used with different value of $\beta$ to sort special sets of OAM modes~\cite{Leach2002, Leach2004, Goyal2021}.

\subsection{Radial mode sorter}

As the name suggests, a radial mode sorter sorts the radial modes instead of $\ell$ modes~\cite{Gu}. Similar to the OAM sorter, the radial mode sorter also uses a Mach-Zehnder interferometer, but instead of using the DPs, it uses a combination of three lenses which introduce radial mode-dependent Gouy phase.

\begin{figure}
    \includegraphics[width=0.35\textwidth]{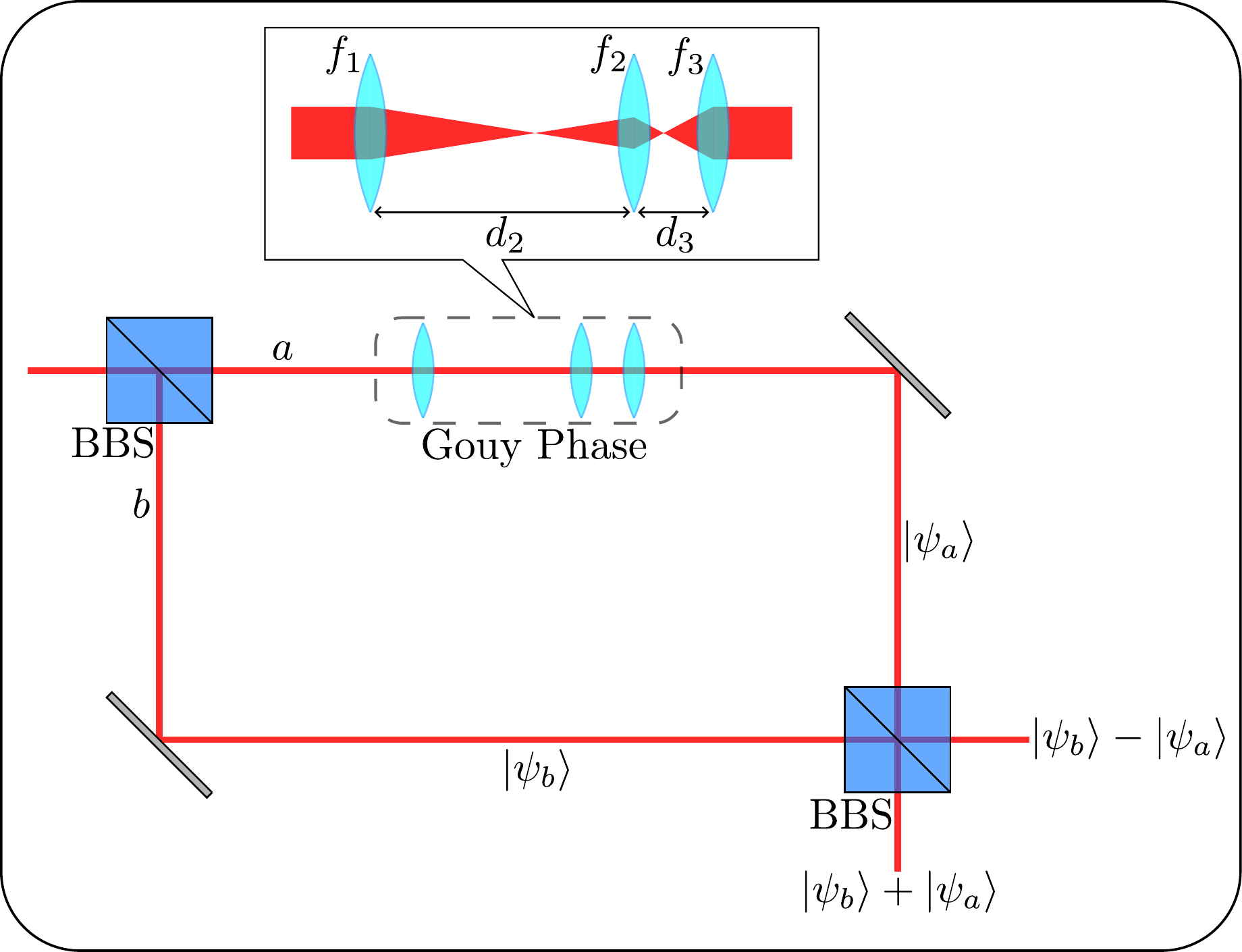}
  \caption{The Gouy-phase radial mode sorter. One arm of the Mach-Zehnder interferometer contains three lenses of focal length $f_1$, $f_2$ and $f_3$ which lead to a relative Gouy phase between the beams in the two arms.}
  \label{Radial_mode_sorter}
\end{figure} 

The radial mode sorter consists of a Mach-Zehnder interferometric setup with a lens configuration or Gouy Phase setup in one of the arms as shown in Fig.~\ref{Radial_mode_sorter}. The lenses cause the state in that arm to accumulate a Gouy phase $(|\ell|+2p+1)\alpha$ relative to the other arm. Here $p$ is the radial index of the LG mode. 
The focal lengths of the lenses $f_1,\:f_2$ and $f_3$, and the distances between them $\:d_1$ and $d_2$ are the parameters that determine the phase $\alpha$, and are chosen such that the beam waist of the beams in the two arms remain the same. Following the analogy with the OAM sorter one can see that this setup can sort the LG modes based on the radial index $p$ for a fixed value of $\ell$. 

\subsection{Auxilliary Hilbert Space Tomography}\label{QST}
A number of schemes to perform partial QST have been proposed over the years~\cite{Jack2009,Nicolas2014,Agnew2011}. Among them, one of the scheme which promises the most efficient and scalable QST for OAM states is the AHST~\cite{Liu2019}. This is a single shot method to perform the QST using the intensity distribution $I(r,\phi)$ of the light beam containing the OAM states.

Let the state of the beam is $\rho$.  The intensity profile measured for this state is given by:
\begin{equation}
\begin{split}
\label{intensity}
I(r,\,\phi)&=\bra{r,\,\phi}\rho\ket{r,\,\phi}\\
			 &=A\sum_{\ell_{1},\,\ell_{2}}f_{\ell_{1}}(r,\,\phi)f_{\ell_{2}}^{*}(r,\,\phi)\rho_{\ell_{1},\ell_{2}},
\end{split}
\end{equation}
where 
$\rho_{\ell_{1},\ell_{2}}=\bra{\ell_1}\rho\ket{\ell_2}$ are the matrix elements of $\rho$ in the OAM basis and $f_{\ell}(r,\,\phi)=\braket{r,\,\phi|\ell}$ is the OAM mode in the polar coordinates.

AHST method exploits the fact that the Fourier transform of $f_{\ell_{1}}(r,\,\phi)f_{\ell_{2}}^{*}(r,\,\phi)$, i.e., $P_{\ell_{1},\,\ell_{2}}(r_{f},\,\phi_{f})=\mathcal{F}[f_{\ell_{1}}(r,\,\phi)f_{\ell_{2}}^{*}(r,\,\phi)]$ satisfies the orthogonality relation
\begin{align}
\iint P_{\ell_{1},\,\ell_{2}}P^{*}_{\ell_{1}',\,\ell_{2}'}e^{\frac{\pi^{2} r_f^{2}w_o^2}{2}}r_fdr_fd\phi_f
=\frac{2}{\pi w_o^2}\delta_{\ell_{1}\ell_{1}'}\delta_{\ell_{2}\ell_{2}'},
\end{align}
for $\ell_1, \ell_2, \ell_1', \ell_2'\geq 0$. 
 $w_o$ is the beam waist at the plane of measurement. Using the orthogonality property we can calculate all the elements of the density matrix $\rho_{\ell_{1},\ell_{2}}$  as
\begin{multline}
\rho_{\ell_{1},\ell_{2}}=\frac{\pi wo^2}{2}\int_{0}^{\infty}\int_{0}^{2\pi}\mathcal{F}[I(r,\,\phi)]P^{*}_{\ell_{1},\,\ell_{2}}\\\times e^{\frac{\pi^{2} r_f^{2}w_o^2}{2}}r_fdr_fd\phi_f.
\end{multline}
We are thus able to perform QST on any state prepared in the subspace spanned by $\{\ket{\ell},\:\ell\geq0\}$.

We can also perform QST on any state prepared in the basis $\ket{\ell},\:\ell\leq0$ using the exact same method because
\begin{equation}
\braket{r,\,\phi|\ell_1}\braket{\ell_2|r,\,\phi}=\braket{r,\,\phi|-\ell_2}\braket{-\ell_1|r,\,\phi}
\label{f_l}
\end{equation}
\begin{equation}
\implies P_{\ell_{1},\,\ell_{2}}=P_{-\ell_{2},\,-\ell_{1}}.
\end{equation}
For any state prepared in the basis $\ket{\ell},\:\ell\leq0$ we can obtain the density matrix from the intensity profile using:
\begin{multline}
\rho_{-\ell_{1},-\ell_{2}}=\frac{\pi wo^2}{2}\int_{0}^{\infty}\int_{0}^{2\pi}\mathcal{F}[I(r,\,\phi)]P^{*}_{\ell_{2},\,\ell_{1}}\\\times e^{\frac{\pi^{2} r_f^{2}w_o^2}{2}}r_fdr_fd\phi_f.
\end{multline}

Unfortunately, AHST can not be used for states prepared in the full basis $\ket{\ell}$ i.e., if the beam contains both positive and negative modes. It is because from \ref{f_l} the terms $\ket{\ell_{1}}\bra{\ell_{2}}$ and $\ket{-\ell_{2}}\bra{-\ell_{1}}$ have the same intensity profile. 

\section{Full state tomography in OAM space of light}\label{Sec:Results}
Although, the AHST scheme for QST works perfectly, the only limitation of this scheme is that it works only for a fixed helicity. Therefore, if the optical beam under consideration has positive as well as negative modes then we can not use this scheme. In this section, we present the scheme to perform full QST for an arbitrary mixed state of OAM modes.  In order to do so, we require a helicity sorter setup which can separate the positive and negative $\ell$ modes. In the following, we show how combining the helicity sorter with AHST method enables us to perform the full QST on the entire OAM space.

Similar to the OAM sorter and the radial mode sorter, the helicity sorter can also be though of as a Mach-Zehnder interferometer with two input and two output modes. Let $a$ and $b$ are the two arms of the interferometer. Let $P_\pm$ are the projectors on the $\pm$ subspaces of the OAM such that $P_\pm = \sum_\ell \ket{\pm\ell}\bra{\pm\ell}~ \forall~ \ell\ge 0$. Then formally the action $S$ of the helicity sorter can be written as
\begin{align}
  \begin{split}
    S = \ket{a}\bra{a} \otimes P_+& + \ket{a}\bra{b} \otimes P_-\\
    &+\ket{b}\bra{a} \otimes P_- +\ket{b}\bra{b} \otimes P_+.
  \end{split}
\end{align}

If $\rho$ is the density matrix of the light beam prepared in the basis $\{\ket{\ell}\}$ with $\ell\neq0$, then we can write it as
\begin{align}
  \rho = \begin{pmatrix}
    \rho_{+} & \sigma\\
    \sigma^\dagger & \rho_{-}
  \end{pmatrix},
\end{align}
where $\rho_{-}$ and $\rho_{+}$ are the marginal density matrices in the negative and positive subspace, respectively, and $\sigma$ and $\sigma^\dagger$ are the cross terms between these two subspaces. Here, the OAM basis is ordered in the following way $\ket{1},\,\ket{2}...\ket{n},\,\ket{-1},\,\ket{-2}...\ket{-n}$ for the ease of calculations. If initially, the beam was in the path $a$ of the helicity sorter, then after passing through the heicity sorter the state reads:
\begin{align}
  \rho_f &= S(\ket{a}\bra{a}\otimes \rho) S^\dagger\\
 &= \ket{a}\bra{a} \otimes \rho_+ + \ket{a}\bra{b} \otimes \sigma +\ket{b}\bra{a} \otimes \sigma^\dagger +\ket{b}\bra{b} \otimes \rho_-.
\end{align}

Performing AHST measurements in the paths $a$ and $b$ will yield the marginal density matrices $\rho_\pm$. However, the cross terms $\sigma$ will be lost in this procedure. In order to estimate the terms $\sigma$ we need to perform $H_x$ and $H_y$ operations on  the density matrix $\rho$,  which are given by
\begin{align}
  H_x &=\frac{1}{\sqrt{2}}
        \begin{pmatrix}
          \mathds{1} & \mathds{1}\\
          \mathds{1} & -\mathds{1}
        \end{pmatrix},\quad
  H_y = \frac{1}{\sqrt{2}}\begin{pmatrix}
          \mathds{1} & i\mathds{1}\\
          i\mathds{1} & \mathds{1}
        \end{pmatrix},
\label{Hx}
\end{align}
where  $\mathds{1}$ is the identity matrix. $H_x$ and $H_y$ perform SU($2$) transformations on all pairs $\ket{\pm \ell},\:\ell>0$ simultaneously. % Application of these operators followed by the helicity-sorter and the measurement in each of the paths will yield the full information about the density matrix $\rho$. Hence, using helicity-sorter and the operators $H_x$ and $H_y$ on the OAM states, one can perform the full QST of OAM states. 

\begin{figure}[h]
  \centering
    \includegraphics[width=0.35\textwidth]{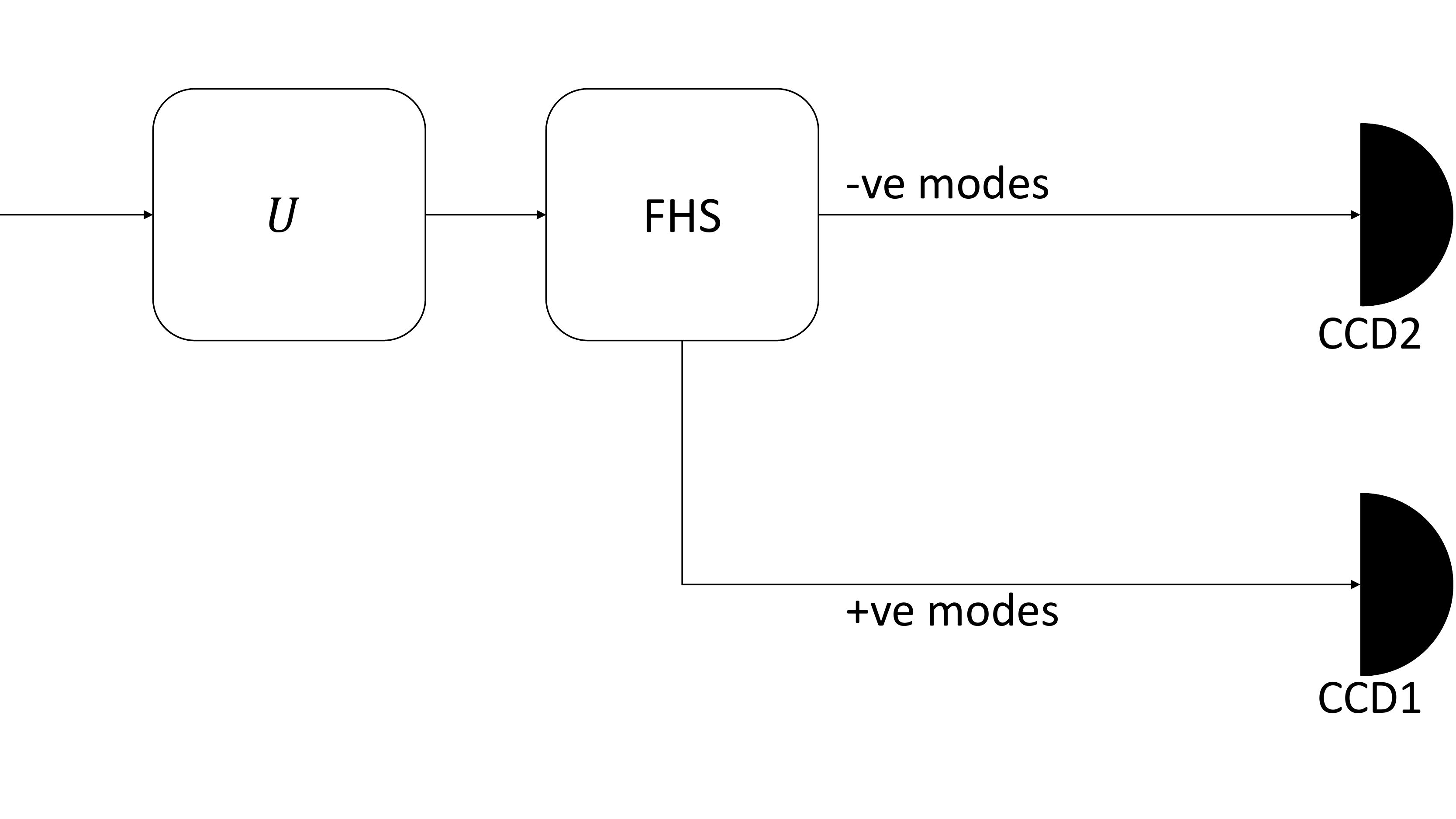}
  \caption{The scheme to perform full QST. Here the operator $U$ representing the measurement setting can take three forms: $U=\mathds{1}$, (ii) $U=H_x$, and (iii) $U=H_y$. The full helicity sorter (FHS) sorts the $\pm$ subspaces of the OAM modes. AHST method is used to estimate the marginal density operators for both subspaces. Combining the results for the three measurement setting one can estimate the full density matrix of the OAM modes. } 
  \label{fig:State_tomography}
\end{figure}

Once we have all the ingredients, we can use the setup shown in Fig.~\ref{fig:State_tomography} to perform the full QST. % The density matrix $\rho$ will be obtained from the intensity profiles measured by the two CCDs.
In order to perform full QST, we need to have three different measurements settings corresponding to (i) $U=\mathds{1}$, (ii) $U=H_x$, and (iii) $U=H_y$. Let marginal states measured by CCD1 are $\mu_1$, $\mu_2$ and $\mu_3$ for the three measurement settings, respectively. Similarly, the marginal states measured at CCD2 are $\nu_1$, $\nu_2$ and $\nu_3$. Therefore,
\begin{equation}
\begin{gathered}
\mu_1=\hat{P}_+\rho\hat{P}_+,\quad \nu_1=\hat{P}_-\rho\hat{P}_-,\\
\mu_2=\hat{P}_+(H_x\rho H_x)\hat{P}_+,\quad \nu_2=\hat{P}_-(H_x\rho H_x)\hat{P}_-,\\
\mu_3=\hat{P}_+(H_y\rho H_y)\hat{P}_+,\quad\nu_3=\hat{P}_-(H_y\rho H_y)\hat{P}_-.
\end{gathered}
\end{equation}
From these six marginal density matrices we can obtain the full density matrix $\rho$ as follows:
\begin{equation}
\rho_+=\mu_1, \quad \rho_-=\nu_1,
\end{equation}
\begin{multline}
\bra{\ell_1}\sigma\ket{-\ell_2}=\bra{\ell_1}\left(\mu_2-i\mu_3-\frac{1-i}{2}\mu_1\right)\ket{\ell_2}\\-\frac{1-i}{2}\bra{-\ell_1}\nu_1\ket{-\ell_2}.
\end{multline}
Here $\ell_1$ and $\ell_2$ are taken to be positive to avoid the repetition. Hence, one can perform full QST on OAM states of light using a helicity sorter and AHST method.
In the following, we present two schemes to implement the helicity-sorter on OAM states of light, and the implementation schemes for $H_x$ and $H_y$ operations.

\subsection{Helicity-sorter}\label{Helicity}

\begin{figure}[h]
  \centering
    \includegraphics[width=0.35\textwidth]{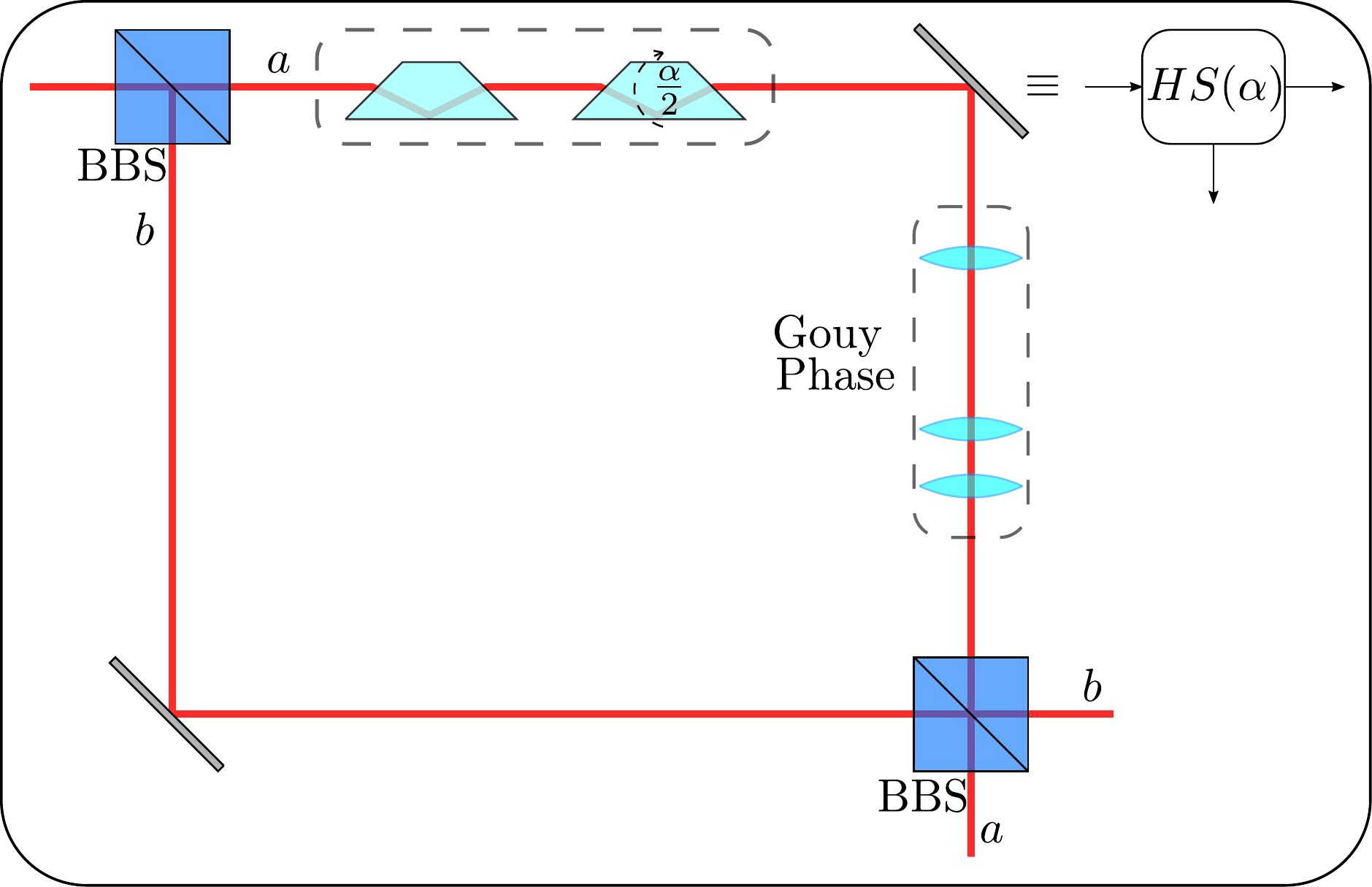}
  \caption{The implementation of a helicity-sorter. BBS stands for Balanced Beam Splitter. The Dove prisms are at a relative angle $\frac{\alpha}{2}$ which rotates the beam by $\alpha$. The parameters of the Gouy Phase setup are chosen such that the beam accumulates a Gouy phase of $\alpha$ relative to the other arm.}
  \label{Helicity_sorter}
\end{figure} 

 The helicity-sorter is a hybrid of the OAM sorter and the  radial mode sorter (see Fig.~\ref{Helicity_sorter}). This setup consists of a Mach-Zehnder interferometer with one arm containing the OAM sorter and the radial mode sorter.   Let the initial state of the beam entering the port $a$ of the BBS is $\ket{\psi}$ which is written as
\begin{align}
  \ket{a}\otimes \ket{\psi} = \ket{a}\otimes\left(\sum_{\ell} c_\ell \ket{\ell}\right),
\end{align}
where $c_\ell$ are the expansion coefficients and the sum is over all the OAM modes from $-\infty$ to $\infty$. Here we are working with OAM modes with $p=0$.

The action of the BBS is given by $\ket{a} \to (\ket{a}+\ket{b})/\sqrt{2}$ and $\ket{b} \to (\ket{a}-\ket{b})/\sqrt{2}$. Upon passing through the beam splitter the state of the beam transforms as
\begin{align}
  \ket{a}\otimes\ket{\psi}\to \frac{1}{\sqrt{2}}\left(\ket{a}+\ket{b}\right) \otimes \ket{\psi}.
\end{align}
In path $a$ we have the Gouy phase setup followed by the dove prism setup that rotates the beam by an angle $\alpha$. The $\ket{\ell}$ mode in path $a$ transforms to $e^{i(|\ell|\alpha+\ell\alpha)}\ket{\ell}$. Therefore, the state after passing through these two arrangements can be written as
\begin{align}
  \ket{\psi_1} = \frac{1}{\sqrt{2}}\left(\ket{a}\otimes \sum_{\ell} e^{i(|\ell|\alpha+\ell\alpha)}c_\ell \ket{\ell}+\ket{b}\otimes \sum_{\ell}c_\ell \ket{\ell}\right).
\end{align}
Upon passing through the second beamsplitter, the state transforms to $\ket{\psi_2}$  which is given by
\begin{align}
  \begin{split}
    \ket{\psi_2} =&   \frac{1}{2}\left(\ket{a}\otimes \sum_{\ell} [1+e^{i(|\ell|\alpha+\ell\alpha)}] c_\ell \ket{\ell}\right.\\
      &+\left.\ket{b}\otimes \sum_{\ell} [1 - e^{i(|\ell|\alpha+\ell\alpha)}]c_\ell \ket{\ell}\right).
  \end{split}\label{Eq:HS1}
\end{align}
One can immediately see that for the negative values of $\ell$ the coefficient of $\ket{b}$ become zero, hence, path $b$ does not support negative OAM modes irrespective of the choice of $\alpha$. However, the positive modes can be in both, path $a$ as well as in path $b$.

We can see that by choosing different values of $\alpha$, different set of positive OAM modes will be present in path $b$. For example, choosing $\alpha = 0$ will result in all positive and negative modes to be in path $a$. Any other value of $\alpha$ will yield some positive OAM modes in path $b$ and other remaining OAM modes in $a$ along with all the negative modes. Therefore, we can call this setup as a partial helicity sorter and represent it by $HS(\alpha)$. In order to realize a full helicity sorter (FHS) we treat even and odd OAM modes separately.

{\em The helicity sorter for odd modes:} We have already seen from Eq.~\eqref{Eq:HS1} that all the OAM modes for $\ell < 0$ are in path $a$. Now choosing $\alpha = \pi/2$ will result in all the odd positive modes being sorted into path $b$. The helicity sorter for odd modes ($HS_o$) can be simply implemented using a partial helicity sorter with $\alpha=\pi/2$: $HS_o=HS(\pi/2)$.

{\em The helicity sorter for even modes:} The helicity sorter for even modes requires additional consideration. One way of constructing the helicity sorter for even modes ($HS_e$) is by using SLMs. An SLM is a device that can be used to apply a phase mask on any OAM mode. Using SLM one can add or subtract one unit of OAM which can cause the transformation $\ket{\ell}\rightarrow\ket{\ell\pm1}$ ~\cite{li2020,BEIJERSBERGEN1994}. Hence, an SLM can be used to convert even modes to odd modes by adding one unit of OAM. The odd modes can be easily sorted using the setup in Fig.~\ref{Helicity_sorter}. Once they are sorted, we can subtract one unit of OAM using an SLM and retrieve the even modes back.

\begin{figure}[h]
  \centering
    \includegraphics[width=0.4\textwidth]{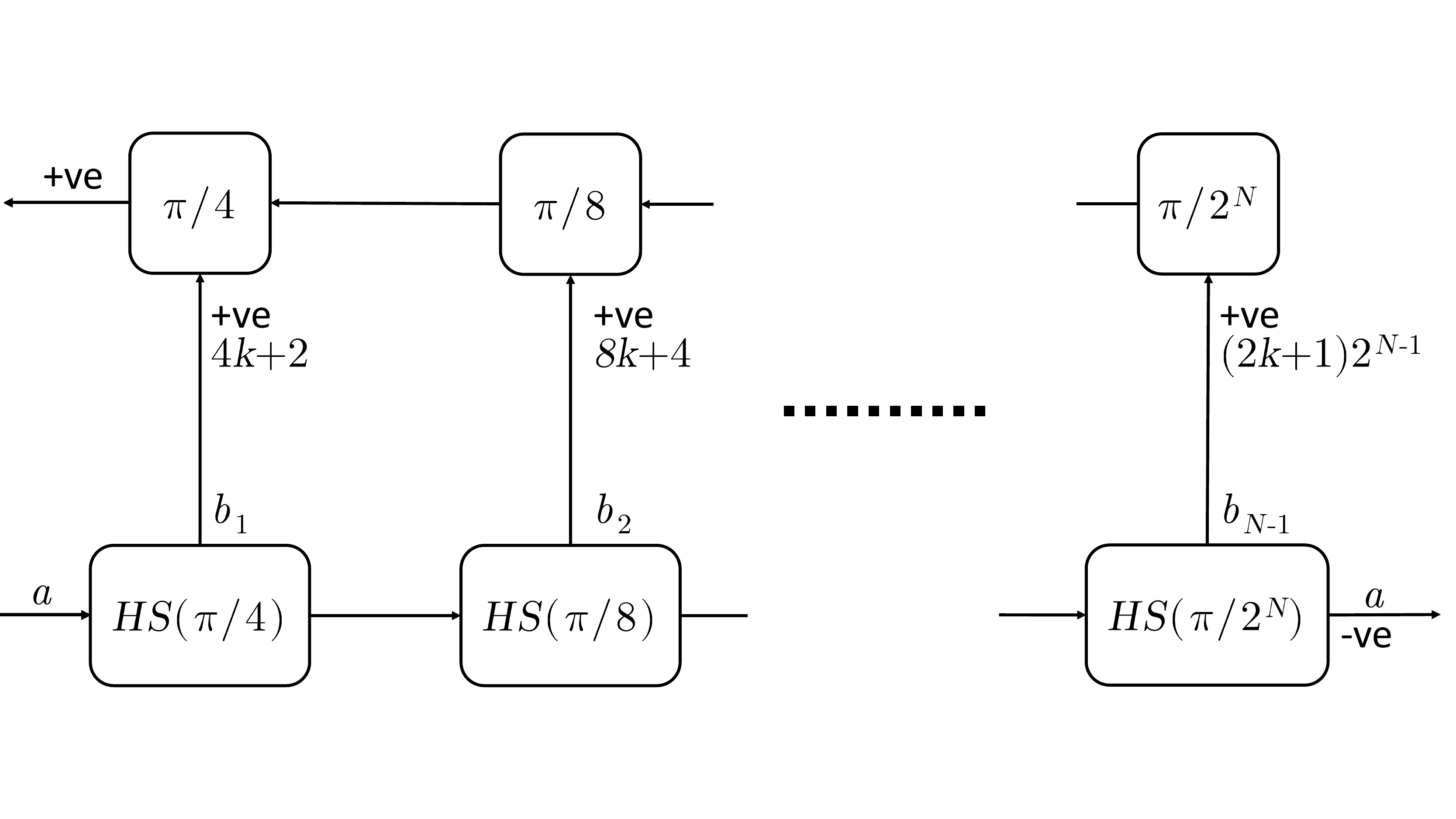}
  \caption{ The implementation of the helicity sorter for even modes $HS_e$ using $N-1$ partial helicity sorter and $N-1$ OAM sorters.}
  \label{FHS}
\end{figure}

Another method to realize $HS_e$ is by using $N-1$ number of partial helicity sorters where the largest value of topological charge in the given OAM state is $\ell_{max}<2^N$. The FHS $HS_e$ can be implemented in two steps:
\begin{enumerate}
	\item We first filter out all the positive modes. In order to do so, we pass the light beam that contains only even OAM modes, through a partial helicity sorter with $\alpha=\pi/4$, i.e., $HS(\pi/4$. Using Eq.~\eqref{Eq:HS1} we can see that $HS(\pi/4)$ filters out all the positive modes  $\ell = 4k+2$ to path $b_1$ (Fig. \ref{FHS}) and all the negative modes and all the positive modes of the form $\ell = 4k$ in path $a$ as shown in Fig.~\ref{FHS}. We then pass the output of path $a$ through  $HS(\alpha=\pi/8)$ to separate all the positive modes of the form $8k+4$ from all the other modes. Continuing in this fashion, we need to pass the beam through a total of $N-1$ partial helicity sorters with parameters $\alpha=\frac{\pi}{4},\,\frac{\pi}{8},...,\,\frac{\pi}{2^N}$, to filter out all the positive modes, for $\ell_{max}<2^N$. At the output of $HS(\pi/2^N)$ we are left with only the negative modes in path $a$.
	\item In order to recombine all the even positive modes scattered across $N-1$ paths: $b_1$, $b_2$,...$b_{N-1}$ (Fig. \ref{FHS}),  we use $N-1$ OAM sorters with $\beta=\frac{\pi}{4},\,\frac{\pi}{8},...,\,\frac{\pi}{2^N}$. The OAM sorter with $\beta=\frac{\pi}{4}$ recombines modes of the form $4k+2$ with $4k$~\eqref{OAMsorter} i.e., the modes in path $b_1$ with the output of the OAM sorter with $\beta=\frac{\pi}{8}$. The OAM sorter with $\beta=\frac{\pi}{8}$ recombines modes of the form $8k+4$ with $8k$ i.e., the modes in path $b_2$ with the output of the OAM sorter with $\beta=\frac{\pi}{16}$ and so on. We get all the positive modes in the same path at the output port of the OAM sorter  with $\beta=\pi/4$.
        \end{enumerate}

This gives us the helicity sorter for even modes $HS_e$~Fig.~\ref{FHS}. It is composed of $N-1$ partial helicity-sorters and $N-1$ OAM sorters. 

\begin{figure}
    \includegraphics[width=0.4\textwidth]{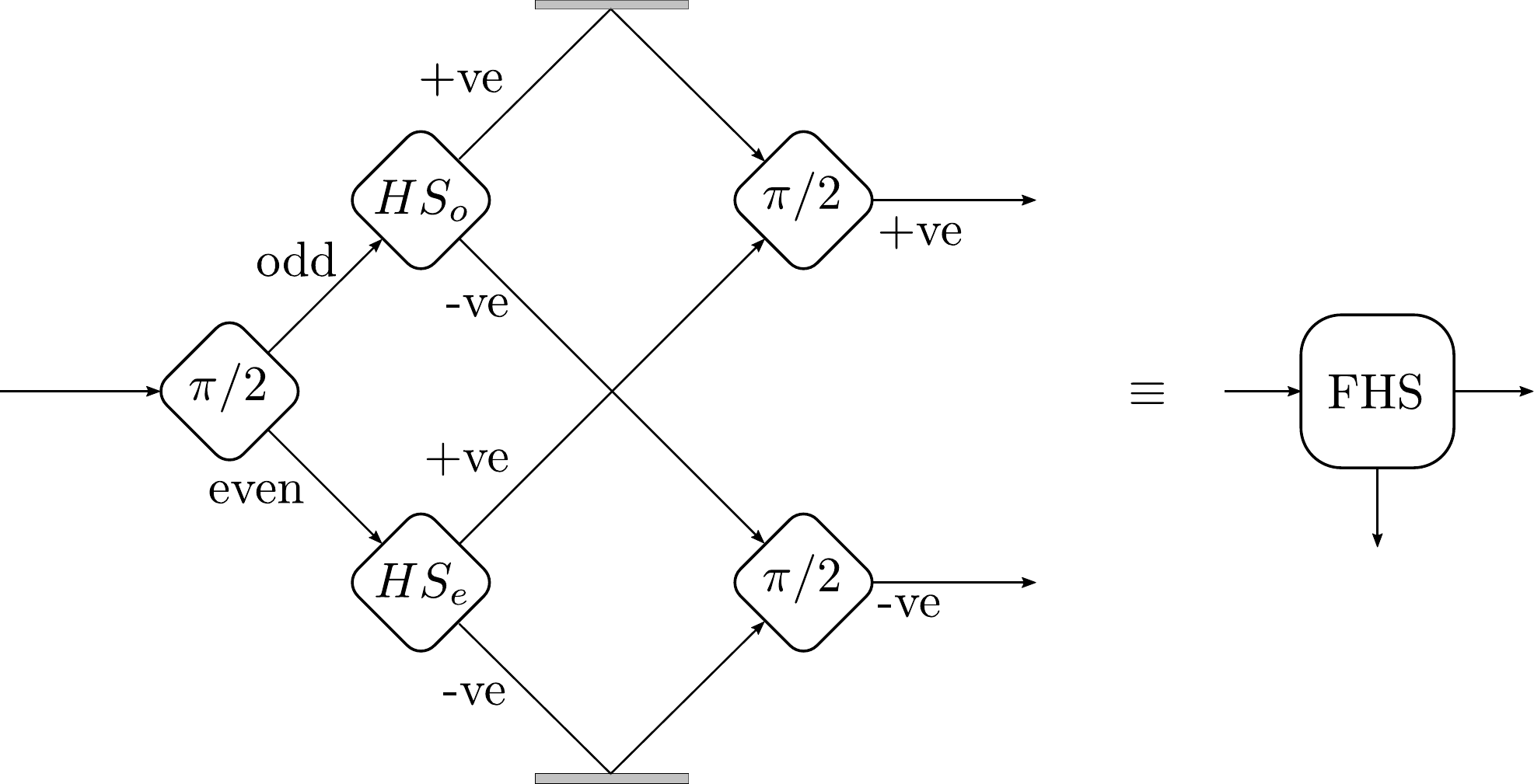}
  \caption{ Schematic diagram for FHS.}
  \label{FHS2}
\end{figure}

Hence, we can realize the helicity sorter for even and odd modes separately. In order to achieve the FHS we first sort the even and odd modes followed by separate helicity sorters on even and odd modes (Fig.~\ref{FHS2}). We then recombine the even positive and odd positive modes together, and the even negative and odd negative modes together using separate OAM sorters (Fig.~\ref{FHS2}). Next we present the implementation scheme for $H_x$ and $H_y$ operators.

\subsection{Implementation scheme for $H_x$ and $H_y$}
\begin{figure}[!ht]
    \subfloat[\label{Unitary1}]{%
      \includegraphics[width=0.4\textwidth]{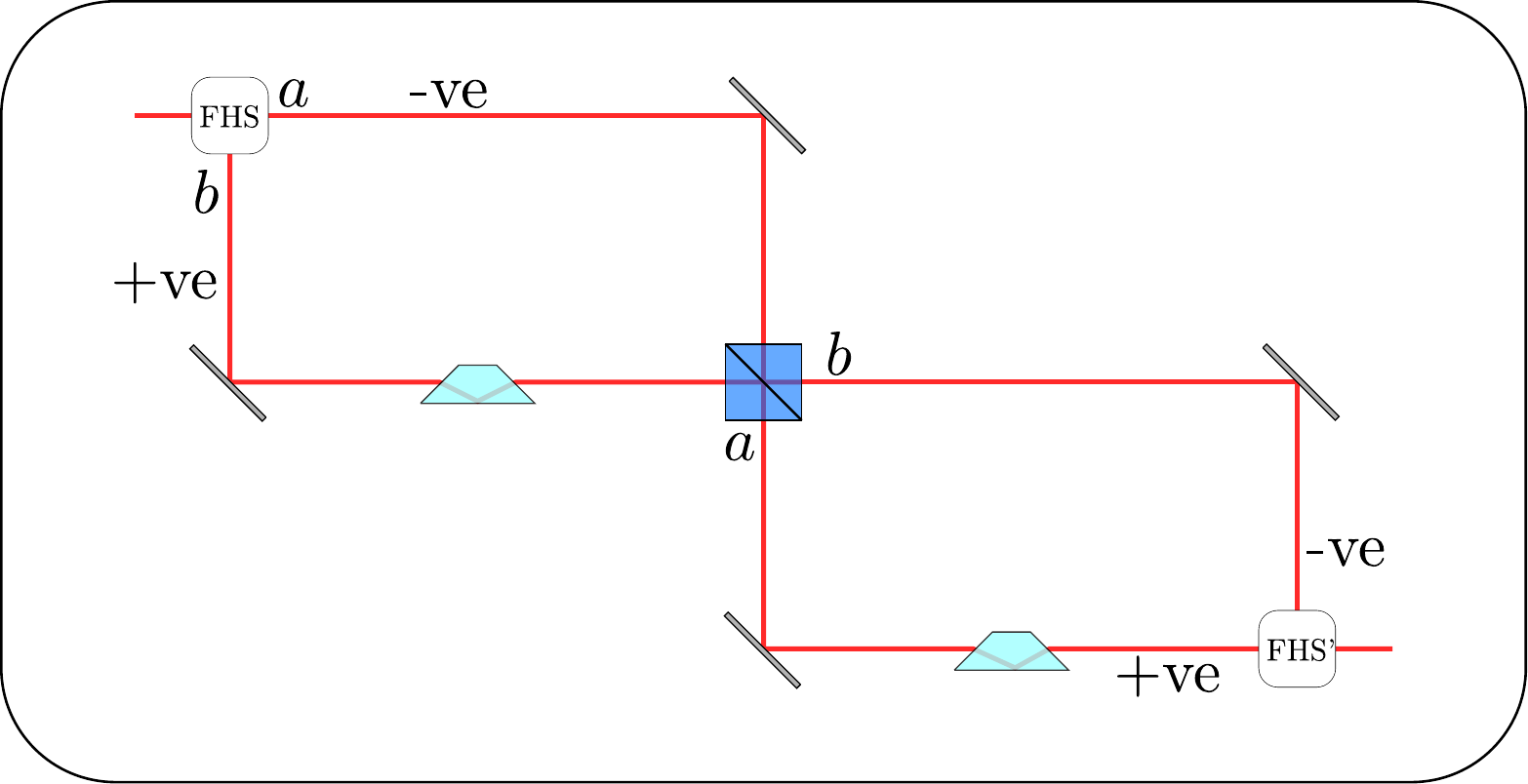}
    }
    \hfill
    \subfloat[\label{Unitary2}]{%
      \includegraphics[width=0.4\textwidth]{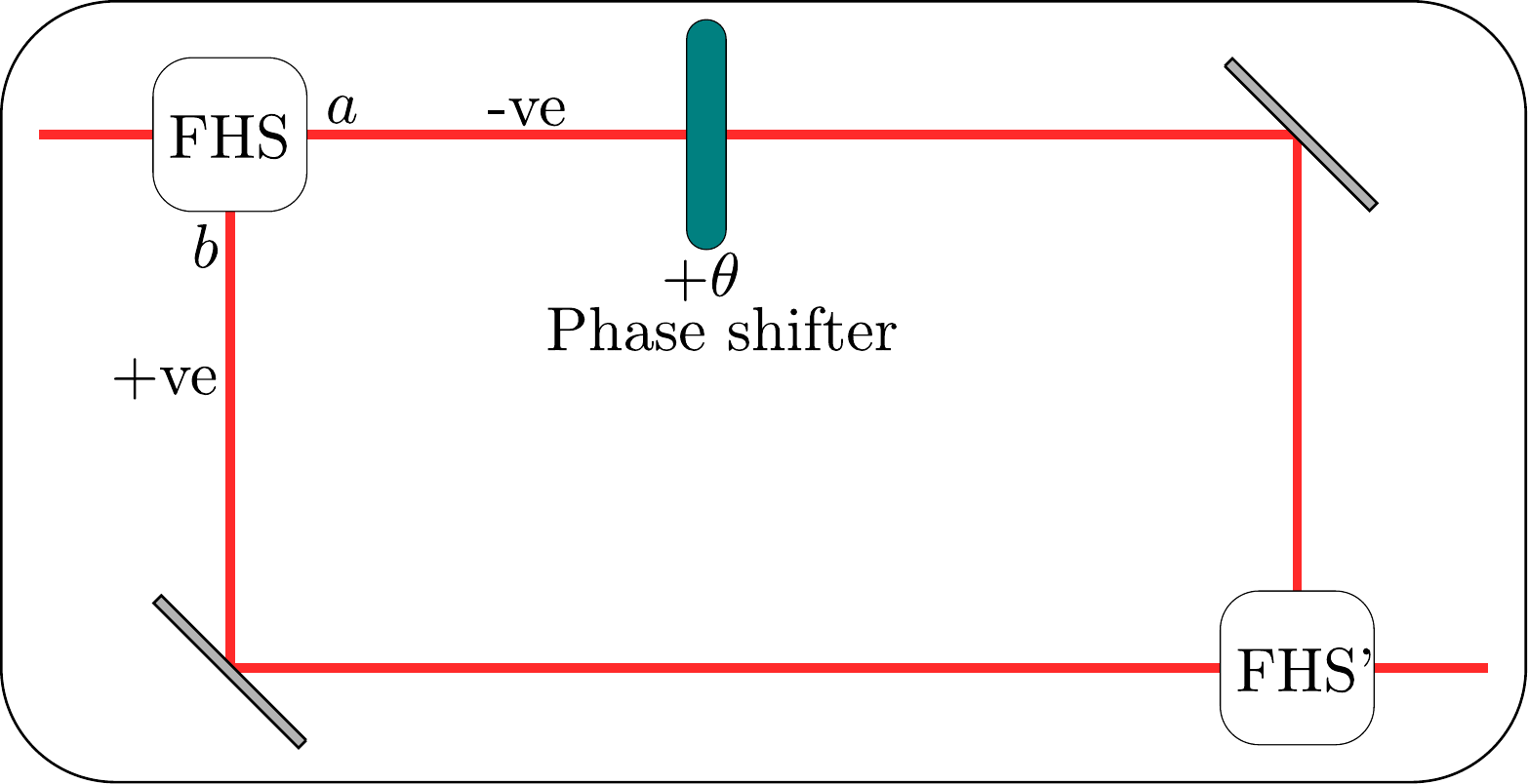}
    }
    \hfill
    \subfloat[\label{Unitary3}]{%
      \includegraphics[width=0.4\textwidth]{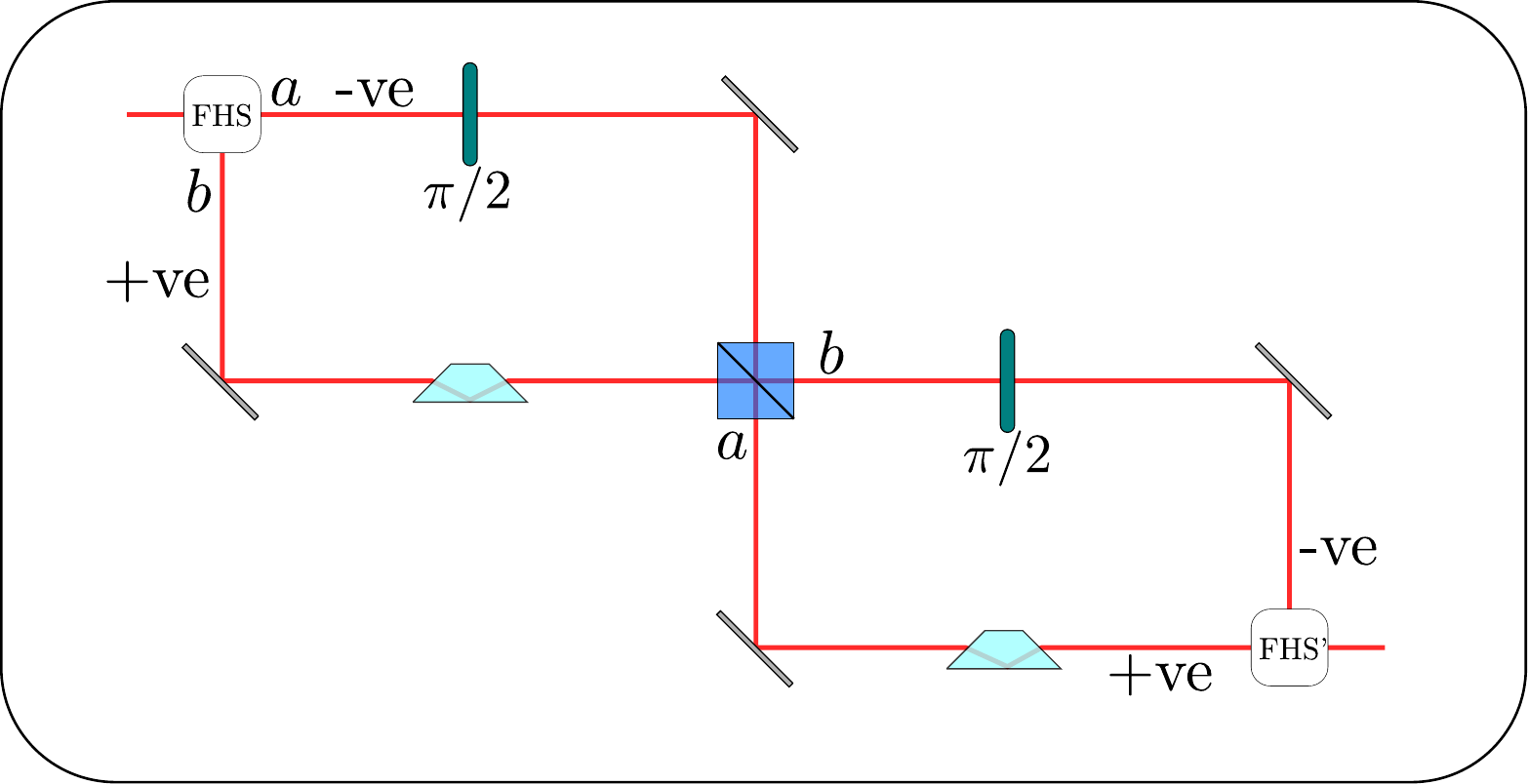}
    }
    \caption{a) The schematics for $H_x$ operator. b) The implementation of the phase gate $P(\theta)$. c) The schematics for $H_y$ operator.}
    \label{Unitary}
\end{figure}

The  $H_x$ \eqref{Hx}  is a Hadamard operator acting on all the pairs of states $\ket{\pm \ell}$ simultaneously. From Eq.~$\eqref{Hx}$, the action of $H_x$ on the OAM states is given by

\begin{multline}
H_{x}\ket{\ell}=\begin{cases} 
      				\frac{1}{\sqrt{2}}(\ket{\ell}+\ket{-\ell})&\ell>0\\
     				\frac{1}{\sqrt{2}}(\ket{\ell}-\ket{-\ell})&\ell<0\\
  		   \end{cases}\:.
\end{multline}
 It can be implemented using the FHS as shown in Fig.~\ref{Unitary1}. Let the initial state in path $a$ is
\begin{align}
  \ket{a}\otimes \ket{\psi} = \ket{a}\otimes\left(\sum_{\ell} c_\ell \ket{\ell}\right).
\end{align}
The action of FHS on this state yields
\begin{equation}
\ket{\psi_1}=\ket{a}\otimes\sum_{\ell}c_{-\ell}\ket{-\ell}+\ket{b}\otimes\sum_{\ell}c_{\ell}\ket{\ell},\:\ell>0.
\end{equation}
The dove prism in path b converts the positive modes to negative modes giving: $\ket{\psi_1'}=\ket{a}\otimes\sum c_{-\ell}\ket{-\ell}+\ket{b}\otimes\sum c_{\ell}\ket{-\ell}$, thus we have negative modes in both paths. The BBS then mixes the modes in the two paths and the second dove prism converts the modes in path $a$ to positive modes, giving:
\begin{multline}
\ket{\psi_2}=\frac{1}{\sqrt{2}}\left(\ket{a}\otimes\sum_{\ell}\left(c_{\ell}+c_{-\ell}\right)\ket{\ell}\right. \\ \left. +\ket{b}\otimes\sum_{\ell}\left(c_{\ell}-c_{-\ell}\right)\ket{-\ell}\right).
\end{multline}

When we recombine the two beams using the FHS we get the Hadamard transformed state (for $\ell>0$):
\begin{multline}
\ket{\psi_3}=\frac{1}{\sqrt{2}}\left(\sum_{\ell}\left(c_{\ell}+c_{-\ell}\right)\ket{\ell}+\sum_{\ell}\left(c_{\ell}-c_{-\ell}\right)\ket{-\ell}\right).
\end{multline}
 The action of $H_y$ on the OAM states is given by~\eqref{Hx}:
\begin{multline}
H_{y}\ket{\ell}=\begin{cases} 
      				\frac{1}{\sqrt{2}}(\ket{\ell}+i\ket{-\ell})&\ell>0\\
     				\frac{1}{\sqrt{2}}(i\ket{\ell}+\ket{-\ell})&\ell<0\\
  		   \end{cases}.
\end{multline}
The transformation $H_y$ can be implemented by combining the transformation $H_x$ with a phase gate $P(\theta)$ and can be written as
\begin{equation}
H_y=P(\pi/2)H_xP(\pi/2).
\end{equation}
The phase gate $P(\theta)$ selectively phase shifts only the negative modes by a phase $\theta$:
\begin{equation}
P(\theta)=\begin{pmatrix}\mathds{1}&0\\
							  0&e^{i\theta}\mathds{1}
			\end{pmatrix}.
\end{equation}
The phase gate can be implemented using the FHS as shown in Fig.~\ref{Unitary2} by applying the phase shift of $\theta$ only on the negative modes. Combining the $H_x$ setup with two of the phase gate setups will result in $H_y$ setup. Since, two FHS applied in series result in identity transformation, the full setup for the $H_y$ looks almost like the one for $H_x$ as can be seen from Fig.~\ref{Unitary3}.

% On close inspection, we see that the action of two FHSs one after the other cancel each other out and give the identity transformation on the modes. Implementing $H_y$ as $P(\pi/2)H_xP(\pi/2)$ is wasteful as it involves two pairs of FHSs occuring one after the other, so 4 of the 6 FHSs used cancel each other out. A more efficient way to implement the $H_y$ transformation is given in Fig.~\ref{Unitary3} after removing the FHSs that cancel each other out. It needs only 2 FHSs.

\section{Conclusion}\label{Sec:Conclusion}
In conclusion, we have presented a method to perform full state tomography for OAM states of a light beam. We have devised method to perform helicity sorter and specific unitary transformations on all the $\pm \ell$ OAM modes simultaneously, which along with AHST method of partial state tomography is used to achieve the task. We have proposed two methods to perform the helicity sorter, one using the SLMs and other using linear optical devices. The number of optical components in scheme with linear optical devices scales as $\mathcal{O}( N)$ for sorting $2^N$ OAM modes, whereas with SLMs, the same setup can be used to sort any arbitrary number of OAM modes. Since  helicity sorter, unitary transformations, and the AHST method, all require only linear optical components, the proposed scheme for full QST is scalable and efficient. This is the first scheme to perform full QST on OAM modes and can be important for quantum computation and quantum communication using OAM modes.

\begin{acknowledgments}
S.K.G. acknowledges the financial support from Inter-disciplinary Cyber Physical Systems (ICPS) programme of the Department of Science and Technology, India, (Grant No.:DST/ICPS/QuST/Theme-1/2019/12).

\end{acknowledgments}

% \bibliographystyle{unsrt}
% \bibliography{bibs}

\end{document}